# Collective Phase-like Mode and the Role of Lattice Distortions at $T_N \sim T_C$ in $RMn_2O_5$ (R= Pr, Sm, Gd, Tb, Bi)

Néstor E. Massa[*],[1] Ali F. García-Flores,[2] Domingos De Sousa Meneses,[3] Leire del Campo,[3] Patrick Echegut,[3] Gilberto F. L. Fabbris,[4] María Jesús Martínez-Lope,[6] and José Antonio Alonso.[6]

[1]Laboratorio Nacional de Investigación y Servicios en Espectroscopía Óptica-Centro CEQUINOR, Universidad Nacional de La Plata, C.C. 962,
1900 La Plata, Argentina,

[2]Instituto de Física "Gleb Wataghin," Universidade Estadual de Campinas, 13083-970 Campinas, São Paulo, Brazil,

[3]CNRS-Conditions Extrêmes et Matériaux Haute Température et Irradiation,
 1D, Av. de la Recherche Scientifique,  F-45071 Orléans, France,

[4]Department of Physics, Washington University in St. Louis, St Louis, MO 63130-4899, USA

[6]Instituto de Ciencia de Materiales de Madrid, CSIC, Cantoblanco, E-28049 Madrid, Spain.

- e-mail : neemmassa@gmail.com

# ABSTRACT

We report on electronic collective excitations in $RMn_2O_5$ (R= Pr, Sm, Gd, Tb) showing condensation starting at and below $\sim T_N \sim T_C \sim$ 40-50 K. Its origin is understood as partial delocalized $e_g$ electron orbitals in the Jahn-Teller distortion of the pyramids dimmer with strong hybridized $Mn^{3+}$-O bonds. Our local probes, Raman, infrared, and X-ray absorption, back the conclusion by which there is no structural phase transition at $T_N \sim T_C$. Ferroelectricity is magnetically assisted by electron localization triggering lattice polarizability by unscreening.

We have also found phonon hardening as the rare earth is sequentially replaced. This is understood as consequence of lanthanide contraction. It is suggested that partially f-electron screened Rare Earth nuclei might be introducing a perturbation to $e_g$ electrons prone to delocalize as the superexchange interaction takes place.

# INTRODUCTION

Complex oxides strong fundamental interactions entangling charge, spins, lattice, and orbital distortions conform a proven field of a plethora of cooperative phenomena rich in novel properties. Compounds developing spontaneous electric polarization coexisting with magnetic order fall in this category. Supporting Jahn-Teller (JT) distortions, and localized carriers, that might also order in stripes,[1] they present an attractive panorama for studying basic interactions at an only energy scale. Congruent magnetic and lattice related phenomena provide grounds for a better understanding on the nature of elementary excitations emerging at magnetic-ferroelectric phase transitions for which there are not yet full experimental consensus. [2-6].

Ferroelectric materials have been described in the basic perovskite structure $ABO_3$. Below the Curie temperature Tc, they undergo a structural phase transition by which the lattice center inversion is lost and spontaneous electric polarization – ferroelectricity- sets in.[7] Although, the phase transition may be broadly considered as displacive or order-disorder more recent local structure studies suggest that real compounds share in different degrees properties belonging to both limits[8] A characteristic hysteresis loop develops in the ferroelectric state under an external electric field that mimics the magnetic susceptibility behavior in magnetic ordered systems.[7]

Materials with simultaneous occurrence of electric and magnetic order are called multiferroics. They have two order parameters, spontaneous polarization (antiferroelectric, ferroelectric, ferrielectric), and spontaneous magnetization (antiferromagnetism, ferromagnetism, ferrimagnetism) triggering, by magnetoelectric coupling, one order by the other. Multiferroics may be grouped depending on how close their lattice is to the perovskite structure. There are compounds with Bi or Pb at the A site. They have two valence electrons participating in chemical bonds through (sp)-hybridized states such as $sp^2$ or $sp^3$. These valence electrons in s-orbitals are the "lone pair" electrons. The lone pair is unstable, mixing the $(ns)^2$ ground state and a low lying $(ns)^1(np)^1$ excited state, thus, leading the ions to break the lattice inversion symmetry yielding ferroelectricity as in $BiFeO_3$.[9] In this matarial the two 6s shell electrons of $Bi^{3+}$ and the magnetic $Fe^{3+}$ ions, at the A and B perovskite sites respectively, results in a compound within the framework of "proper" ferroelectrics. On the other hand, those materials in which the electric polarization develops as consequence multiple interactions that may include a pre-set magnetic order, are named improper ferroelectrics. $RMn_2O_5$ (R= Rare Earth, Bi)) belong to this last category. [11]

From annoying impurity minor phases in the preparation of $RMnO_3$ (R=La, Pr, Nd), and isostructural to $YMn_2O_5$, the family of compounds $RMn_2O_5$ (R= R=Rare Earth, Bi) grew to be the source of great research activity due to their sharing of low temperature magnetic and ferroelectric ordering.[12]

$RMn_2O_5$ (R=Rare Earth, Bi) are complex oxides containing manganese ions located at distinct crystallographic oxygen environments in a mixed valence state achieving stabilization of the nominal high spin oxidation states $Mn^{3+}$ and $Mn^{4+}$ [12,13]. They have a lattice built on infinite chains of $Mn^{4+}O_6$ octahedra sharing edges linked by $Mn^{3+}O_5$ pyramid dimmers. Bismuth, and Rare Earth ions are in distorted $RO_8$ polyhedra. where Bi, unlike Pr, Sm, Gd, and Tb, has the two valence electrons in s-orbitals configuring the unstable lone electron pair.[9],[14]. The octahedra-pyramid interconnection generates frustration in the spin structure since not all nearest neighbor Mn-O-Mn superexchange interactions are satisfied.[15]

The motivation of the present work is the lack of agreement in understanding lattice order and the multiferroic nature of $RMn_2O_5$ (R= Rare Earth, Bi). While X-ray diffraction yields centrosymmetric P*bam*, Z = 4, space group from 300 K to 4 K, the appearance of an effective antiferromagnetic order below $T_N$~39-45 K, coexisting with spontaneous lattice polarization (ferroelectricity) few degrees below $T_C$, implies a not observed structural change with loss of the inversion center as in the Pb$2_1$m, subgroup of the space group P*bam*.[15-19] The loss of the inversion center results in the positions for ions Mn2, O2, O3 and O4 split into two kind of sites.[20]

The same arguments apply for phonons independently measured in Raman and infrared spectra. These do not show the expected change due to the hypothetical gerade-ungerade split, meaning infrared phonons becoming also Raman active and vice versa below $T_N$~$T_C$, predicted in the P*bam* space group but absent in P$2_1$am[21,22]. At the transition temperatures, only very few uncorrelated and very weak new vibrational modes are detected at internal mode frequencies. Overall, the available data points for RMn$_2$O$_5$ (R= Rare Earth, Bi) a lattice topology described by a unique space group above and below $T_N$~$T_C$.

We also report on a temperature dependent collective electronic excitation in the 10 cm$^{-1}$ to 90 cm$^{-1}$ frequency range in a temperature driven scenario triggering condensation at $T_N$~$T_C$. In addition, we found an unexpected oscillator mass dependence for vibrational modes suggesting that Rare Earth nuclei may contribute as a perturbation source of the electronic cloud in a charge/orbital density wave scenario.[23]

**EXPERIMENTAL DETAILS**

Our sample preparation follows procedures reported by Alonso et al [14]. Stoichiometric amounts of analytical grade R$_2$O$_3$ and Mn(NO$_3$)$_2$ 4H$_2$O were solved in citric acid. The citrate solutions were slowly evaporated and decomposed at temperatures up to 600ºC. All the organic materials were eliminated in a subsequent treatment at 800ºC in air. The samples were annealed at 1273 K either in a high oxygen pressure atmosphere (200 bar), for (R= Pr, Sm, Gd) or in air for (R= Tb, Bi).[12]. Our X-ray (CuKα) diffraction patterns for $RMn_2O_5$ (R= Bi, Rare Earth) are shown in Fig. 1.

Temperature dependent infrared reflectivity spectra with 2 cm$^{-1}$ resolution were taken with polycrystalline pellets mounted on a cold finger of a home-made cryostat in a FT-IR Bruker 113v interferometer. A gold mirror was used as 100% reference.

Raman scattering spectra were excited with the 514.5 nm laser line from an Ar$^+$ laser, with power of ~12 mW focused in a spot of ~100 μm diameter made in a near back-scattering configuration. The scattered light was analyzed by a Jobin-Yvon triple grating spectrometer equipped with a LN$_2$-cooled CCD detector. The samples were mounted on a cold finger of a closed-cycle He refrigerator. All measurements were taken on heating runs.

We complemented those techniques with transmission- X-ray absorption measurements, done at D04B-XAFS2 beamline in the Brazilian Laboratory for Synchrotron Light (LNLS), adding local structural information on the long range averaging view of X-ray diffraction.[12, 13, 14]

## RESULTS AND DISCUSSION

*i) Spin-phonon interactions*

Early Raman scattering showed the presence of strong phonon modulation of the Mn-O-Mn magnetic superexchange electron correlations well above the magnetic ordering temperature at $T_N$.[22,24]

Raman band splits also hinted of a lower than expected lattice symmetry when the spectra of lighter lanthanide compounds were compared against heavier and Bi counterparts. Figure 2 and 3 shows some new results in consonance with those findings.

The onset of magnetoelastic effect on infrared active vibrational modes below T*~65 K is also shown in Figure 4 where the temperature dependence of the internal stretching and bending phonon bands are found to have a turning point. At ~$T_N$ the oscillator strength Eq. 6, see below) diminishes as the samples cool down. This effect is better defined in $BiMn_2O_5$,[25] and shows that the displacement patterns modulating Mn-O-Mn [26] are common to infrared and Raman spectra. As pointed by Garcia-Flores et al [22, 24] magnetic correlations couple to phonon frequencies through the superexchange integral associated with a modulation of the Mn-O-Mn bonding angles and/or Mn-O stretching vibrations. It is also in agreement with results for $DyMn_2O_5$ reflecting the strong spin-lattice coupling dynamics above the magnetic ordering temperature.[27]

*i)   Debye-Waller factors*

Lower lattice symmetry hints for Rare Earth heavier ions were corroborated by our study on Bi first oxygen shell Debye-Waller factors.

Debye-Waller factors are shell-dependent factors that are most sensitive to modes contributing to radial motions, basically, depending only on the local vibrational structure. We study of nearest neighbors Bismuth-Oxygen pairs conforming the asymmetric polyhedra $BiO_8$ that allows knowledge of the relative motion of atomic pairs in an analysis of extended X-ray absorption fine structure (EXAFS) spectra. The mean square variation in bond lengths is a quantity that has two components, one arising from the thermal vibrations and another from temperature dependent structural disorder.[28]

We worked in the harmonic approximation, a single frequency Einstein approximation, not taking into account contributions from low structural disorder and dynamical effects due to ion motions. Then, the Debye-Waller factor applied to the Bi-O atomic pair reduces to

$$\sigma^2(T) = \sigma_0^2 + \left(\frac{\hbar^2}{2\mu \cdot k_B \theta_E}\right) \cdot \coth\left(\frac{\theta_E}{2T}\right) \qquad (1)$$

Where $\sigma_0^2$ is the static contribution and $\theta_E$ is the so-called Einstein temperature for the ion pair with reduced mass $\mu$. $\theta_E$ is the fitting parameter of the experimental data [28]

Its temperature dependence within that approximation shows no anomalies on distances distribution at about $T_N \sim 39$ K or elsewhere pointing that to achieve an acceptable fit to the experimental data it is only necessary to invoke one space group, in our case, the orthorhombic Pbam.[29] Therefore, within our resolution (~0.01Å), we could not identify any structural transition.

It does, however, establish the existence of two sets of Bi-O bonds. This is shown in Figure 5 for the first Bismuth coordination shell with two Einstein temperatures, $\theta_{E1} = 294 \pm 7$K, for the four ions and $\theta_{E2} = 462 \pm 28$ K for the rest of the eight Bi-O bonds. Moreover, the relative low values for $\theta_E$ also suggest that the polyhedra configuring the Bi-O first shell are not as rigid as Mn sublattices where $\theta_E$ ranges ~700 K.[29]

These results also corroborate earlier measurements by Tyson et al [30] for Tb in TbMn$_2$O$_5$ where it was found an anomalous Tb-O bond distribution resolving

below~180 K into two peaks, denoting unequivalent Tb sites, away from the Pbam structure. This pictures an important departure from the results measured in diffraction patterns since it implies a lower symmetry space group that the fitted Pbam.

## ii) Far Infrared Reflectivity

Figs. 6 and 7 show the near normal infrared reflectivity of $RMn_2O_5$ (R=Pr, Gd, Sm Tb) from 300 K to 4K in the 10 $cm^{-1}$ to 800 $cm^{-1}$ frequency range. Lower frequency vibrational modes below ~250 $cm^{-1}$ correspond to the Rare Earth beating against polyhedra while internal modes, antisymmetric stretching, bending, and symmetric breathing modes are between ~250 $cm^{-1}$ and ~600 $cm^{-1}$ and ~600 $cm^{-1}$ and ~770 $cm^{-1}$ respectively. [27, 31], The spectra seek to emphasize one more time that there is no substantial changes in the phonon profiles denoting a possible structural modification when phonon cooling and hardening take place.

In addition, a featureless band much stronger than any phonon structure appears below 100 $cm^{-1}$. Centered at ~30 $cm^{-1}$, it is weakly temperature dependent down to $~T_N~T_C$, where it shows a smooth transition. We associate its origin to a collective electronic contribution. Its energy position and profile, and the absence of Raman activity in that frequency range, rules out possible contributions of rattling modes.

Figure 8 contrasts the overall spectra of $GdMn_2O_5$ against that of $BiMn_2O_5$ at 300 K and 4K. The smooth low frequency band against the discrete phonon spectrum is highlighted with circles. In $BiMn_2O_5$, the Bi electronic configuration tends to further stabilize the lattice as the stereochemical activity of the Bi 6s lone pairs takes place on the orbitals.[32] This effect reduces that feature to a broad background on which, being infrared spectroscopy an additive technique, phonon profiles are delineated.

The temperature dependent detail of the low frequency reflectivities for $PrMn_2O_5$, $GdMn_2O_5$, $SmMn_2O_5$ and $TbMn_2O_5$ is shown in figure 9. We note, that even if it seems to be rare-earth dependent, the overall behavior is preserved showing lower temperature condensation, and thus, electron localization. At about $T_N$, but at slightly higher temperature, the band becomes less intense and its profile significantly broadens as if a random uncorrelated charge-spin reordering were taking place. A few degrees further below the band hardens as at the onset of a spin pinned charge density wave perhaps also denoting an incipient spin density wave. The small temperature range for the relative weak band condensing below $T_N$ precludes for giving a more quantitative comparison with similar phenomena reported in $R_{1-x}Sr_{1+x}MnO_4$ (R=Nd and La)[33] and for $LaSrMnO_3$ [34]. Nevertheless, and since there is not yet a neutron data counterpart for our compounds, we may extend the discussion in terms of conclusions drawn by Pailhes et al [35] for inelastic neutron scattering measurements in $YMnO_3$. It was found that at low frequencies hybridization takes place due to the mixing of spin waves and phonons. These are well defined zone center features interpreted as multiferroic goldstone modes. Spin gaps were measured at 2.2 and 5 meV [35]. Very recently, we found that those bands match our lower frequency reflectivity for $RMnO_3$ (R=Rare Earth) [36]. Accordingly, here we propose that in $RMn_2O_5$ (R= Pr, Sm, Gd, Tb, Bi), a similar phenomena is taking place at and below $T_N \sim T_C$. This would involve dislocated, d-orbital $e_g$ electrons being infrared active by dynamical multiferroic coupling. The mode emerging below the ordering transitions, shown in the insets of figure 9, corresponds to this picture.

Being $RMn_2O_5$ (R= Pr, Sm, Gd, Tb) insulators the smooth low frequency band may be simulated above $\sim T_N \sim T_C$ by one dipole. A second one was added reproducing the partial condensation below transition temperatures.

The dielectric function, $\varepsilon(\omega)$,[37] above $\sim T_N \sim T_C$ is then given by

$$\varepsilon(\omega) = \varepsilon_1(\omega) - i\varepsilon_2(\omega) = \varepsilon_\infty \frac{\left(\omega_{LO}^2 - \omega^2 + i\gamma_{LO}\omega\right)}{\left(\omega_{TO}^2 - \omega^2 + i\gamma_{TO}\omega\right)} \quad . \quad (2)$$

where $\varepsilon_\infty$ is the high frequency dielectric function; $\omega_{TO}$ and $\omega_{LO=}\omega\text{"phase-like mode"}$, are the transverse and longitudinal optical frequencies, and $\gamma_{TO}$ and $\gamma_{LO}$ are their dampings, respectively.

The dielectric function real ($\varepsilon_1(\omega)$) and imaginary ($\varepsilon_2(\omega)$) part (complex permitivity, $\varepsilon^*(\omega)$) is then estimated with fits to the data collected in reflectivity measurements using the reflectivity R given by.

$$R(\omega) = \left| \frac{\sqrt{\varepsilon^*(\omega)} - 1}{\sqrt{\varepsilon^*(\omega)} + 1} \right|^2 \quad (3)$$

This procedure allows the knowledge of the real part temperature dependent optical conductivity, $\sigma_1(\omega)$, proportional the absorption coefficient, [38]

$$\sigma_1(\omega) = \frac{\omega \cdot \varepsilon_2}{4\pi} \quad , \quad (4)$$

with maxima at transverse optical frequencies.

The detection of an associated field to longitudinal optical frequencies in near normal reflectivity measurements is clue as band peaking in the energy loss function,

$$\text{Im}\left(\frac{1}{\varepsilon_1(\omega)+i\varepsilon_2(\omega)}\right). \quad (5)$$

It is verified by angle dependent reflectivity measurements.[39]

In addition, we also calculated the Sj strength of the j$^{th}$ oscillator (fig 4) as

$$S_j = \omega_{jTO}^{-2} \frac{(\prod_k \omega_{kLO}^2 - \omega_{jTO}^2)}{(\prod_{k \neq j} \omega_{kTO}^2 - \omega_{jTO}^2)} \quad (6)$$

Our results for the optical conductivity and the energy loss of $PrMn_2O_5$ at low frequencies are shown in figure 10. They are representative of all our Rare Earth samples. The optical conductivty temperature dependence (i.e., transverse excitations), Fig. 10(a), has the same behavior as reported in absorption measurements by Sushkov [40] et al and Golovenchits and Sanina (Figure 5)[41] in $RMn_2O_5$ (R=Rare Earth). It also reproduces the same phonon soft mode hardening when the sample is cooled pointed out by Shuvaev et al [42] in simpler manganites.

The energy loss spectrum, figure 10(b), peaking at the longitudinal optical frequency undergoes a small softening and narrowing consequence of the limited number of electrons experience condensation as it goes through the transition temperatures into the antiferromagnetic-ferroelectric phase.

*iii) Rare Earth lattice distortion*

The rare earth vibrational dependence is quantitatively brought up by a remarkable effect at all in the internal mode frequencies. As it is shown in Fig. 11 for symmetric stretching modes, as the number of the electrons in the f shell increases there is an overall hardening of phonon frequencies. This goes against what intuitively may be expected from the classical oscillator mass dependence as found for the corresponding bands in the spectra of $BiMn_2O_5$ also shown in the same figure 11.

We interpret this effect as due to the lanthanide contraction, i.e., as consequence of the partial screening of f-electrons, 6s electrons are attracted toward the nucleus decreasing the ionic radius, and consequently, stiffening bonds making the cage and altering lattice constant systematically.[13] Covalence varies and points to an indirect but first order rare-earth dependence of the Mn-O hybridization as $RO_8$ cage distortions that depend on the rare earth thus triggering an overall perturbation on nearest neighbor orbitals. Being a structural change, it is easily followed by infrared spectroscopy at all temperatures and independently verified by Raman scattering (Fig. 11, inset).

*iv) Discussion*

Our far infrared reflectivities (Fig. 9 ) suggests activated electric dipoles in the meV range of the same nature of those reported by Han et al[43] in the paraelectric phase of $RMn_2O_5$ (R= Tb, Dy, Er) for the complex dielectric permittivity using impedance measurements at 100 Hz to 1 MHz. They concluded that that the thermally activated dipoles were product of an intrinsic bulk process. not involved in magnetic related interactions.

The scenario prompted by our measurements is reminiscent to the collective nature of $e_g$ carriers in colossal magnetoresistance materials and within the context of charge density waves (CDW) already known in doped manganites. In these systems there is definitive manifestation of strong electron-lattice coupling creating polaronic carriers, charge hopping, a dynamic Jahn-Teller effect, and an antiferromagnetic background that localizes charges at $T < T_N$ and may lead to electronic regular patterns.[1, 44] In $La_{1-x}Ca_xMnO_3$ (x=0.5, 0.67) the Bardeen-Cooper-Schrieffer (BCS) relation for an optical gap $2\Delta(T)$ opened at $T=T_0$ seems to hold if a ratio $(2\Delta(T_0)/k_BT_C) \sim 5$ is allowed.[45]

That implies optical detection of a CDW phase mode and a CDW amplitude mode.[46] The Raman active amplitude mode (CDW position dependent), a soft-phonon-like mode [47, 48] and the far infrared active phase mode (CDW time dependence), resulting from an electronic cloud beating against the positive background, are paradigmatic excitations of quasi-one-dimensional metals such as blue bronzes $M_{0.3}MoO_3$ (M=K, Rb).[49]

A BCS-type temperature dependence for the gap was also found by Katsufuji et al [50] in $La_{1.67}Sr_{0.33}NiO_4$. Here the ratio $2\Delta(0)/K_BT_{CO} \sim 13$ is understood as indicative of unaccounted electron-electron interactions or fluctuations arising from low dimensionality. The view is the known for polar manganites, where electrons tend to delocalize in lower dimension layered structures with strong electron-phonon interactions localizing charge leading to lattice polarons [51]

However, having verified the inexistence in $RMn_2O_5$ (R= Rare Earth, Bi) of a low frequency Raman excitation in the temperature range 20 K to 300 K, there is need of a broader approach than the BCS single particle excitation across a gap for discussing collective excitations. Keeping a ratio of ~20 for a gap as in $TbMn_2O_5$ of ~1.4 eV [52] yields an unrealistic high temperature below which might be triggered. a hypothetical electron quasi-ordered pattern

In RMn$_2$O$_5$ (R= Rare Earth) we are only bound by the detection of net giant electric dipole created by a fluctuating electronic cloud alike a BCS-phase mode. The materialization of this electronic cloud interacting as a balance to lattice restoring forces would be the only similarity to the originally proposed by Fröhlich [53] for a collective mode as consequence of the Peierls lattice distortion.[54] On the other hand, as pointed by Blawid and Millis,[55] extremely large electron-phonon interaction can maintain that picture if localized carriers as polarons, with large activation gap, are considered. This results in a large gap to ordering temperature ratio, and is also consistent with no high temperature metal phase being detected above it. A source of polaron formation in our compounds is in the Mn$_2$O$_{10}$ Jahn-Teller dimmer distortions. The collective band (Fig. 9) is concurrent to Mn$^{3+}$ pyramids 3d superexchange interactions and bond-charge fluctuations in a distorted complex Mn$_2$O$_{10}$ where electrons are strongly trapped within Mn sites.[56] This picture also helps to explain the origin of a low energy excitation contribution reported for the near constant thermal expansion coefficient of BiMn$_2$O$_5$ [25] and the large thermal dependence of Tb-O distances in TbMn$_2$O$_5$.[30]

The emerging result is thought as consequence of the competition between the tendency of electrons to delocalize in RMn$_2$O$_5$ (R=rare earth) against increasing the charge-lattice interaction, assisted, below $T_N$, by the underlying antiferromagnetic order [57,58]. This it is also coincident with the remarkable interrelationship between magnetic and electric polarization properties in millimeter wave dielectric constant found by Hur et al.[18]. The magnetic assisted condensation may be the reason on why the antiferromagnetic ordering is found always few degrees above the ferroelectric transition. Mn$^{3+}$ $e_g$ electrons, as it is already known for Mn$^{3+}$ octahedral simpler manganites [44,59] will be linked to Jahn–Teller lattice distortions. Polaron fluctuations and condensation at $T_N$ may be interpreted as originating in a Jahn-Teller coupled charge density wave allowing by hybridization the detection of an spin density wave below $T_N$. Dislocated, d-orbital $e_g$ electrons, that might also be partially spin correlated in the paramagnetic phase, condense in spin wave modes coupled to small atomic displacements yielding spin-lattice hybrid as measured in YMnO$_3$ [35, 59] These excitations are detected in the infrared at low frequencies due to complex dynamical multiferroic couplings. Strong Mn-O hybridization is required for spin configuration inducing charge redistribution [52]. This is also supported by oxygen spin polarization measurements at the onset of ferroelectricity denoting oxygen sites hybridized with Mn$^{3+}$ off-center ions [60]

No substantial changes in crystallographic data are expected when passing $T_N \sim T_C$, and, thus, in the number of phonons detected. The quenched Jahn-Teller orbital in the dimmer will locally unveil uncorrelated very weak new internal vibrational modes at low temperatures that will also correlate to the small shifts in Mn$^{3+}$ lattice position measured by neutron diffraction .[61]. The net weak lattice polarization (ferroelectricity) at $\sim T_C$ would then be a natural consequence.

Summarizing, we report on a far infrared collective phase-like-mode and on the multiferroic phase origin in RMn$_2$O$_5$ (R= Pr, Sm, Gd, Tb, Bi). The pyramid sublattice deformation is concomitant to the absence of the inversion center at the lower temperature unscreened picture allowing detection of weak spontaneous lattice polarization (ferroelectricity). This is due to electron condensation taking place in a system assisted by the magnetic order in an already higher-temperature non-centrosymmetric space group. At $T_N$ and below, there are no substantial changes in crystallographic data described by an X-ray effective Pbam space group, and accordingly, no change in the number of phonons measured across $\sim T_N \sim T_C$. Being RMn$_2$O$_5$ (R= Rare Earth, Bi) structurally close to a layered topology, strong localized and correlated polarons, as suggested by the ~0.4 eV tail in the TbMn$_2$O$_5$ reported optical gap of ~1.4 eV [52] would play a significant role. Conclusions from recent calculations on the structure and ferroelectric polarization of $TbMn_2O_5$ support the view of a charge/orbital ordered ground state driven by the octahedra and pyramids where the main role in the polarization is due to $Mn^{3+}$ ions in the pyramids. It also points to the magnetic structure assisting to the lattice polarization (ferroelectricity) in consonance with our proposition.[62]

Lanthanide contraction causes phonon hardening as the number of f-electron increases and may introduce a systematic orbital perturbation.

## ACKNOWLEDGEMENTS

N.E.M. is grateful to the CNRS-C.E.M.H.T.I. laboratory and staff in Orléans, France, for research and financial support in performing far infrared measurements. N.E.M., G.F.L.F., G.M.A. would like to thank the Brazilian Synchrotron Light Laboratory (Laboratorio Nacional de Luz Sincrotron-LNLS) for economic assistance and making available beam time at the D04B-XAFS1 beamline. L. D. C. thanks the Basque Government for post-doctoral financial support at the CNRS-C.E.M.H.T.I. NEM. also acknowledges partial financial support (PIP 0010) from the Argetinean Research Coouncil (Consejo Nacional de Investigaciones Científicas y Técnicas-CONICET). Funding through Spain Ministry of Science and Innovation (Ministerio de Ciencia e Innovación) under Project Nº MAT2010-16404 is acknowledged by J. A. A. and M. J. M-L.

# REFERENCES

[1]     2012 Resnik, D.  arXiv: 1202.0852

[2]     Goldstone J, Salam A., and  Weinberg S 1962 Phys. Rev **127**  965

[3]     de Souza R and  Moore J. E:  2008 Phys Rev B **77** 012406

[4]     Katsura H,  Balarsky A. V., and Nagaosa, N 2007 Phys. Rev. Lett. **98** 027300.

[5]     Mostovoy J,  2006 Phys. Rev. Lett. 96 067601

[6]     Pimenov A, Shuvaev A. M.,  Mukhin A. A. and Loidt A. 2008 J. Phys.Condens. Matter **20** 434209

[7]     Lines M. E. and  Glass A. M., Principles and Applications of Ferroelectrics and Related Materials, Clarendon Press, Oxford, UK, 1977

[8]     Stern E. A., 2004 Phys. Rev. Lett. **93** 037601

[9]     Wang K.F., Liu J. M., and  Ren Z. F.,  2008 Adv. Phys. **58**, 321

[10]    Massa N. E.,  del Campo L.,  De Sousa Meneses D,  Echegut P, . Frabbris G. F. L.,  Azevedo G.,  Martínez-Lope M-J., and  Alonso J. A. 2010 J. of Appl. Phys. **108** 084114

[11]    Ribeiro J- L-, 2007 Phys. Rev. B **76**  144417

[12]    Alonso J. A., Casais M. T.,  Martinez-Lope M- J., and Rasines J. 1997 Solid State Chem. **129** 105

[13]    Muñoz A., J A. Alonso, M- T. Casais, M. J. Martínez-Lope, J. L. Martínez, and M. T. Fernández-Diaz  2005 Eur. J. Inorg. Chem. **2005** 685

[14]    Alonso J. A., Casais M. T., Martínez-Lope M. J., Martínez J. L. and Fernández-Díaz  M. T.1997 J. Phys.  Condens. Matter **9** 8515

[15]    Chapon L. G,. Radaelli P. G.,  Blake G R.,  Park S., and Cheong S.-W 2006 Phys. Rev. Lett. **96** 097601)

[16]    Muñoz, A., J. A. Alonso,  M. T. Casais, M. J, Martinez_Lope, J. L. Martinez, and M. T. Fernández-Diaz 2002 Phys. Rev. B **65** 144423

[17]     Kobayashi S., Osawa T., Kimura H., . Noda Y,. Kagomiya I, and Kohn K. 2004 J. Phys. Soc. Jpn. **73**, 1593

[18]    Hur N.,  Park S-,  Sharma P. A.,  Guha S., and Cheong S.-W.  2004 Phys.Rev. Lett. **93** 107207

[19]    Kagomiya I., S. Matsumoto, K. Kohn, Y. Fukuda, T. Shoubu, H. Kimura, Y. Noda, and N. Ikeda 2006 Ferroelectrics **286**, 167

[20]    Volkova L. M. and  Marinin D. V. 2009 J. Phys.: Condens. Matter **21** 5903

[21]     Mihalova B., Gospodinov M. M.,  Guthler B.,  Yen F,  Liyvinchuk A. P. and Live M. N. 2005 Phys. Rev. B **71** 172301

[22]    García-Flores A. F.,  Granado E., Martinho H., Urbano R. R., Rettori C., Golovenchits E. I.,  Sanina V. A., Oseroff S. R., Park S., and Cheong S.-W 2004 Phys. Rev. B **73**, 104411

[23]    Efremov  D. V.,, Van den Brink J., Khomskii D I.2004 Nature Materials **3** 853.

[24]    García-Flores A. F.,  Granado E., Martinho H., Urbano R. R., Rettori C., Golovenchits E. I.,  Sanina V. A., Oseroff S. R., Park S., and Cheong S.-W 2007 J. Appl. Phys. **101** 09M106.

[25]    Granado E., Eleotério M. S., García-Flores A. F.,  Sopuiza J. A., Golovenchits E. I. and Sanina V. A. 2008 Phys. Rev. B **77** 134101

[26]    Litvinchuk A. J., 2009  J. of Magnetism and Magnetic Materials **312**, 2373

[27]    Cao J.,  Vergara L. I, Musfeldt  J.L.,  Litvinchuk A. J.,  Wang Y- J.,  Park S., and Cheong S.-W 2008  Phys. Rev. B **78** 064307

[28]    X-Ray Absorption: Principles, Applications, Techniques of EXAFS, SEXAFS and XANES, Edts. D.C. Koningsberger and R. Prins, John Willey and Sons, New York, (1988).

[29]    Fabbris, G. F. L.  MSc Thesis, Institute of  Physics, Universidade Estadual de Campinas, Campinas, Brazil, (2009).

[30]    Tyson T-, M. Deleon, S. Yoong, and S. -W. Cheong 2007  Phys. Rev. B **75**, 174413

[31]    Valdés Aguilar R,  Sushkov A. B.,  Park S., Cheong S.-W, and. Drew H. D. 2006 Phys. Rev. B **74** 184404

[32]     Seshadri R, and Hill N. A., 2001 Chem Mater. **13** 2892

[33]    Fujioka J-, Y. Ida, Y. Takahashi N Kida, R. Shimano and Y- Tukura 2010 Phys Rev. B **82**, 140409R.

[34]    Nucara A., Maselli, P. Calvani, R. Sopracase, M. Ortolani, G. Gruener, M-Cestelli Guidi, U. Schade, and J. Garcia 2008 Phys. Rev. Lett. **101** 066407.

[35]     Pailhès S,  Fabrèges X, .Régnault L. P., Pinsard-Godart L,  Mircheau I,, Moussa F.,  Hennion M and  Petit S, 2009 Phys. Rev B **79** 134409

[36]     Massa N. E.,  del Campo L,  De Sousa Meneses D.,   Echegut P., Martínez-Lope  M. J. , and Alonso J. A., 2012 Bull. of Am. Phys. Soc, **17,** Nº1, http://meetings.aps.org/link/BAPS.2012.MAR.T32.14.

[37]     Kurosawa  T 1961 J. Phys. Soc. Jpn **16**, 1298

[38]    Wooten F., "Optical properties of solids", Academic Press (San Diego), (1972).

[39]     Berreman D. W. 1963 Phys. Rev. **130**, 2193. (See also,  Massa N. E., Denardin J. C., Socolovsky . M.,  Knobel M., and  Zhang X. X.,2009  J. of  Appl. Phys. **105**, 114306).

[40]    Sushkov A. B., Valdés Aguilar R.,  Park S., Cheong S.-W, and Drew H. D.2007  Phys. Rev. Lett. **98** 027202

[41]     Golovenchits E. and  Sanina Y. 2004 J. Phys. Condens. Matter **16** 4325.

[42]   Shuvaev A. M., . Hemberg J., Niemann D., Schrette F., Loidl A., Yu V., Ivanov J., Travkin V. ., Mukhin A. A.and Pimenov A. 2010 Phys. Rev. B **52** 174417.

[43]    Han T, C., Lin J. G., Kuo K. M., and G. Chem 2008 J. Appl. Physics **103** 084106.

[44]   Tokura Y. 2006 Rep. Prog. Phys. **69** 797

[45]   Calvani P., de Marzi G., DoreP., Lupi S., Maselli P., d'Amore F. and Gagliardi S, 1998 Phys. Rev. Lett. **81** 4504.

[46]   Gruner G. 1988 Rev. Mod. Phys. **60** 1129.

[47]   Travaglini G., Mörke I., and Wachter P. 1983 Solid State Commun. **45** 289.

[48]    Massa N. E.1990 Solid. State, Commun. **76**, 805.

[49]   Travaglini G. and Wachter P. 1984 Phys. Rev. B **30** 1971.

[50]   .Katsufuji T., Tanabe T., Ishikawa T., Fkuda Y., Arima T., and Tokura Y. 1996 Phys. Rev. B **54** 14230.

[51]    Zaanen J. and Littlewood P.B. 1994 Phys. Rev B **50** 7222

[52]    Moskin S.and Pisarev R. V. 2008 Phys. Rev. B **77** 060102(R)

[53]   Fröhlich H. 1954 Proc. R. Soc. London Ser. A **223** 269

[54]   Peierls R. E., Quantum Theory of Solids, (Clarendon, Oxford) 1955.

[55]   Blawid S.and Millis A. (2000) Phys. Rev. B **62** 2424

[56]   Daoud-Aladine A., J. Rodrigues-Carvajal, L. Pinsard_Gaudart, M. T. Fernández_Díaz, and A. Revcolevschi, Phys. Rev. Lett. **89**, 097205 (2002).

[57]    Golovenchits E. I, Morozov . E. V., Sanina V- A., and Sapozhnikova L. M. 1992 Sov. Phys. Solid State **34**, 56

[58]   Campbell B. J., Osborn R., Argyriou D. N., Vasilu-Doloc I., Mitchell J. F., Sinha S. K., Ruett U., Ling C. D., Island Z., and Lynn J. W- 2001 Phys. Rev. B **65**, 014427

[59]   Sato J,  Lee S. H.,  Katsufuji T,  Mesaki M-,  Park S-, . Copely J. R. D.,  and Takagi H 2003 Phys. Rev. B, **68** 014432)

[60]   Beale T- A- W-,  Wilkins S. B., Johnson R. D.,  Bland S. R.,  Joly Y., . Forrest T. R., McMorrow D. F.,  Yakhou F.,  Prabhakaran D.,  Boothroyd A. T., and  Hatton P. D.. 2010 Phys. Rev. Lett. **105** 087203 105.

[61]   Blake G. R.,  Chapon L. C.,  Radaelli P. G.,  Park S.,  Hur N., Cheong V, and Rodriguez-Carvajal J. 2003 Phys. Rev. B **71** 214402

[62]   Tay-Roy Chang, Horng-Tay Jeng, Chung-Yuan Ren, and Chen-Shiung Hsue 2011  Phys. Rev. B **84** 024421

# FIGURE CAPTIONS

**Figure 1** *(color online)* X-ray (CuKα) diffraction patterns for $RMn_2O_5$ (R= Bi, Rare Earth),

**Figure 2** *(color online)* Raman spectra of the ~200 cm$^{-1}$ vibrational band in $RMn_2O_5$ (R=Pr, Eu, Bi) at 300 K and 12 K . It suggests that replacements of lighter by heavier ions yields a lower space group symmetry even at 300 K. Note that in every case there is no change in the number of bands measured between 300 K and 12 K;

**Figure 3** *(color online)* Temperature dependent Raman active $PrMn_2O_5$ phonons showing the effects of strong phonon modulation on the Mn-O-Mn magnetic superexchange electron correlations well above the magnetic ordering temperature at $T_N$.

**Figure 4** *(color online)* Vibrational bands for internal modes of $PrMn_2O_5$ and $BiMn_2O_5$ as a function of temperature showing the effect of magnetism in changes of the relative intensities (oscillator strengths) at and above the ordering temperatures. The continuous line on the top of the bands is a representation of temperature dependence of the calculated oscillator strength- (eq. 6, see text).

**Figure 5** *(color online)* Temperature dependent Debye-Waller factors for $BiMn_2O_5$ showing two non-equivalent bond distances for Bi nearest neighbors in the $BiO_8$ sublattice cage. Right panel shows the two sets of Bi-O bonds in the $BiMn_2O_5$ lattice.

**Figure 6** *(color online)* Temperature dependent near normal reflectivity of RMn2O5 (R=Pr, Sm) shown in increments of 10 K from 4 to 80 K and in increments of 20 K from 80 to 300 K. Rectangles highlight the low frequency phase-like mode. For better viewing the spectra have been displaced vertically.

**Figure 7** *(color online)* Temperature dependent near normal reflectivity of RMn2O5 (R= Gd, Tb) shown in increments of 10 K from 4 to 80 K and in increments of 20 K from 80 to 300 K. Rectangles highlight the low frequency phase-like mode. For better viewing the spectra have been displaced vertically.

**Figure 8** *(color online)* Near normal infrared reflectivity of $GdMn_2O_5$ and $BiMn_2O_5$ at 4 K and 300 K. Circles highlight the low frequency phase-like mode smooth profile against the phonon discrete in $BiMn_2O_5$.

**Figure 9** *(color online)* Near normal reflectivity detail of the temperature dependent phase-like mode in $RMn_2O_5$ (R=Pr, Sm, Gd, Tb). The insets show the same spectra vertically displaced. To assure reproducibility some of these measurements were also thermally cycled around ~$T_N$-$T_C$.

**Figure 10** *(color online)* (a) Optical conductivity and (b) energy loss function for the phase-like mode in $PrMn_2O_5$. The inset shows the optical conductivity displaced vertically.

**Figure 11** *(color online)* Frequency shifts due to lanthanide contraction in $RMn_2O_5$ (R=Pr, Sm, Gd, Tb) of the symmetric stretching modes at 300 K and 4 K. Inset: Raman spectra of the phonon Rare Earth dependence at 12 K. For comparison the band corresponding to $BiMn_2O_5$ is also shown in both set of measurements.

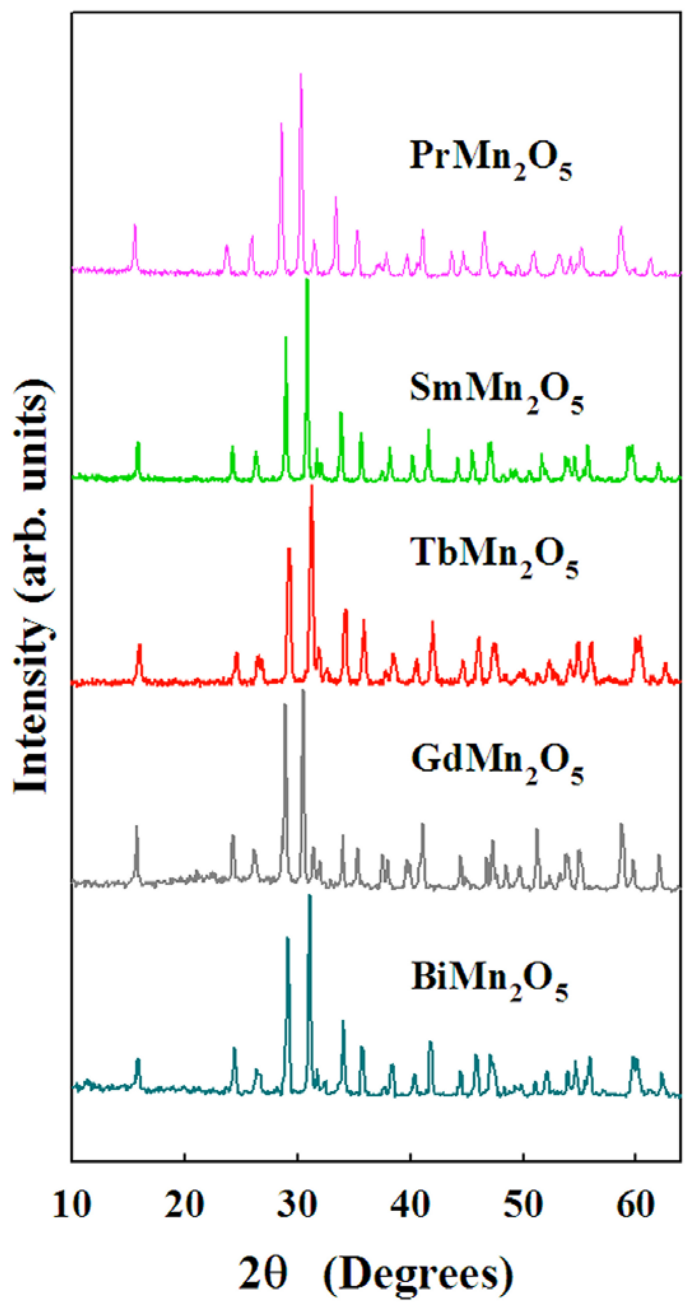

Figure 1
Massa et al

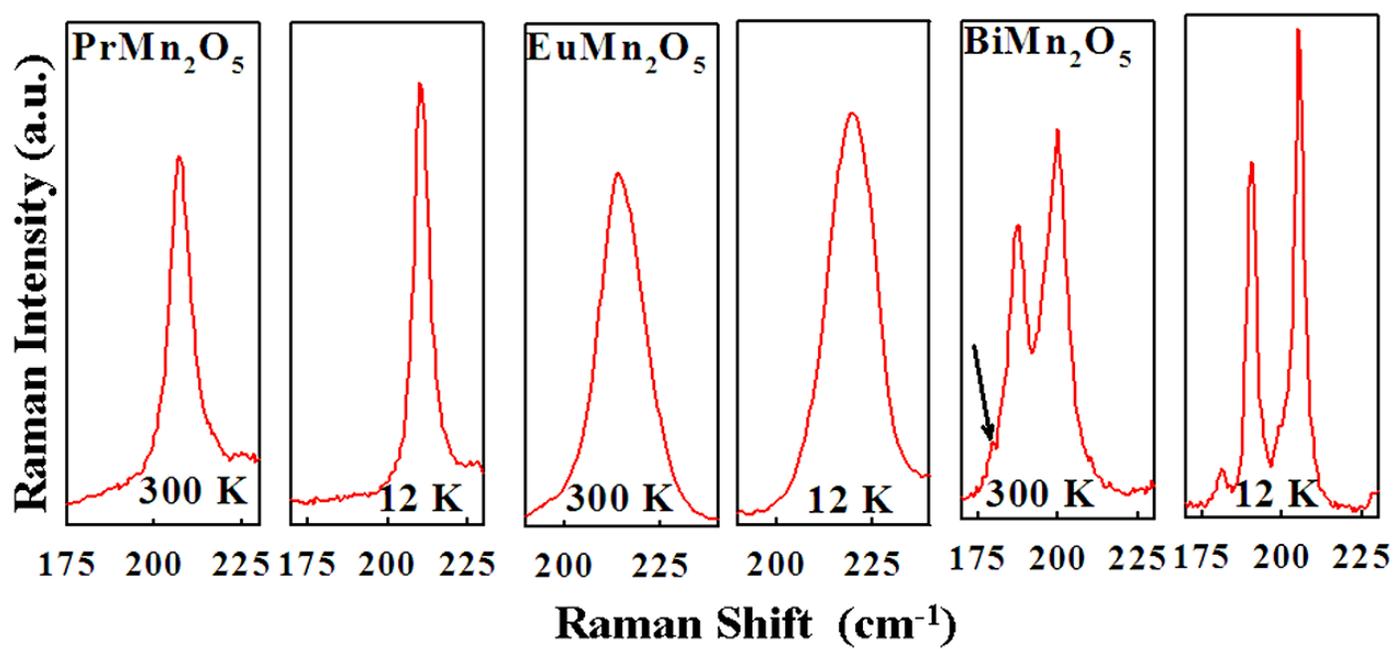

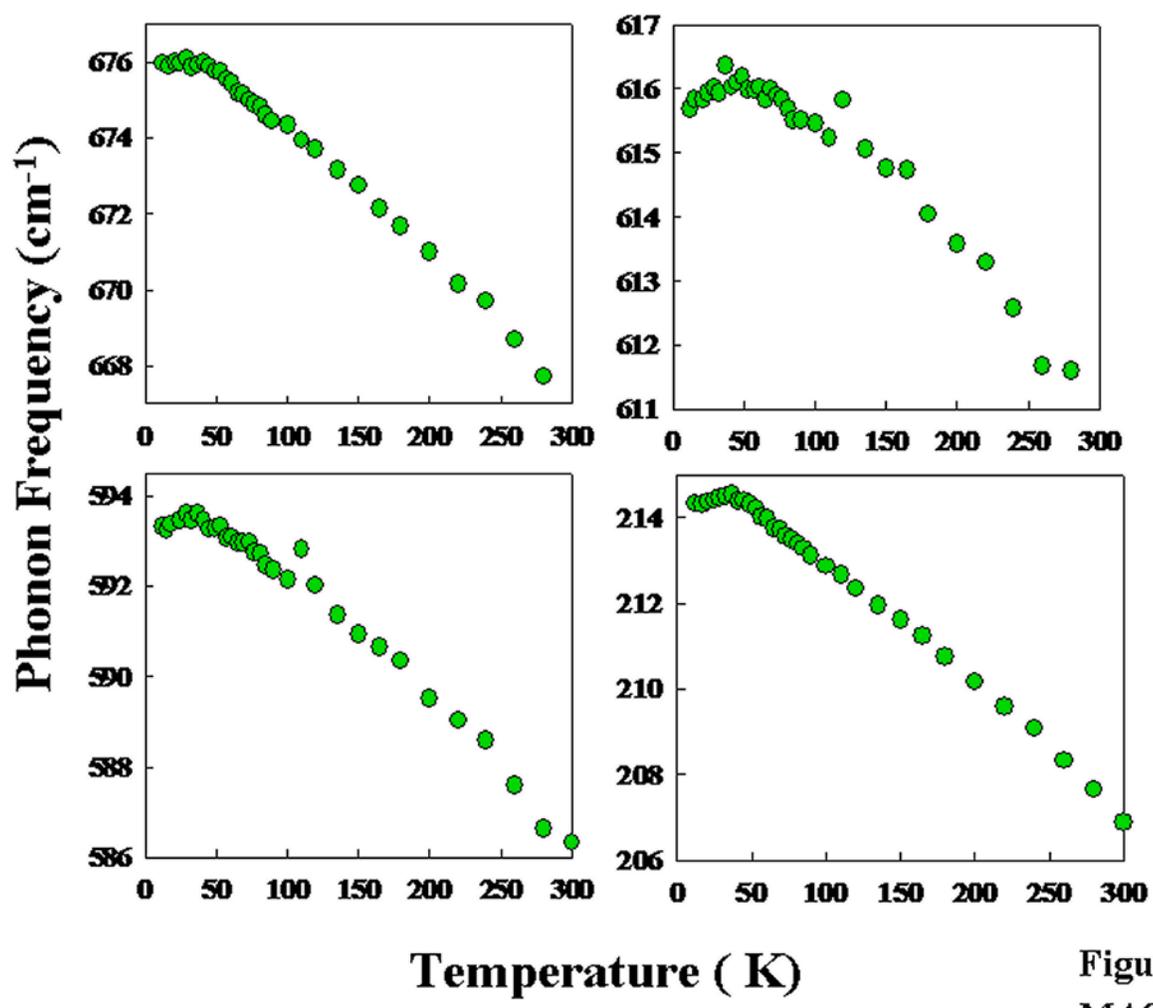

Figure 3
MASSA et al

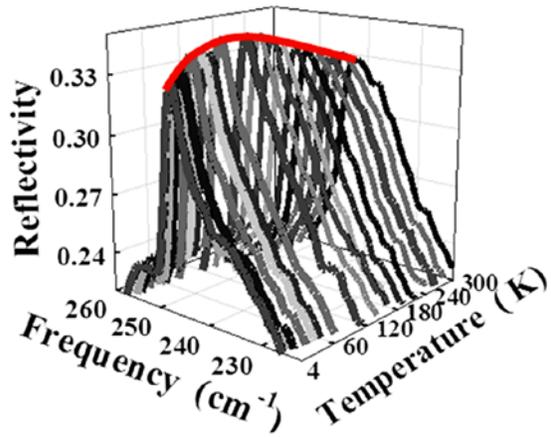
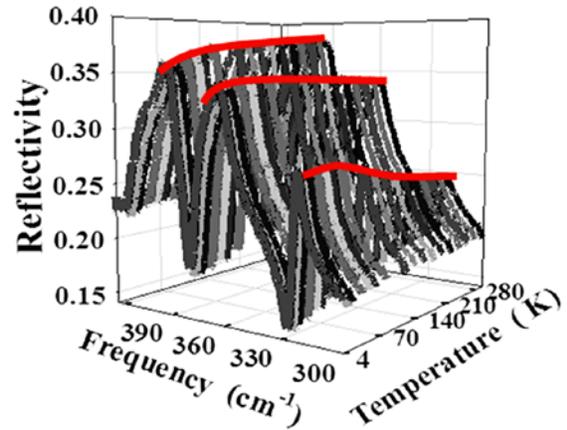
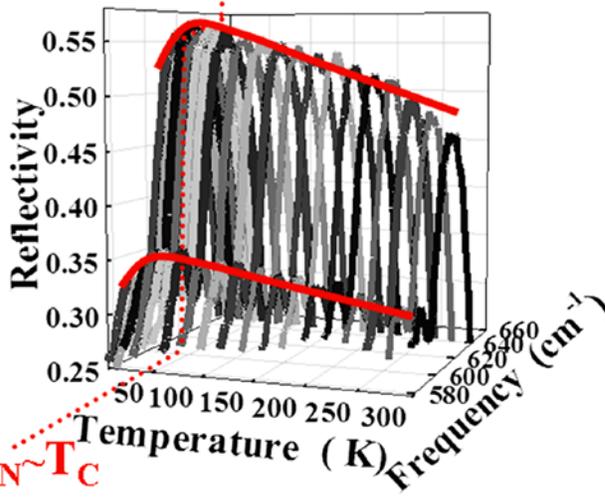
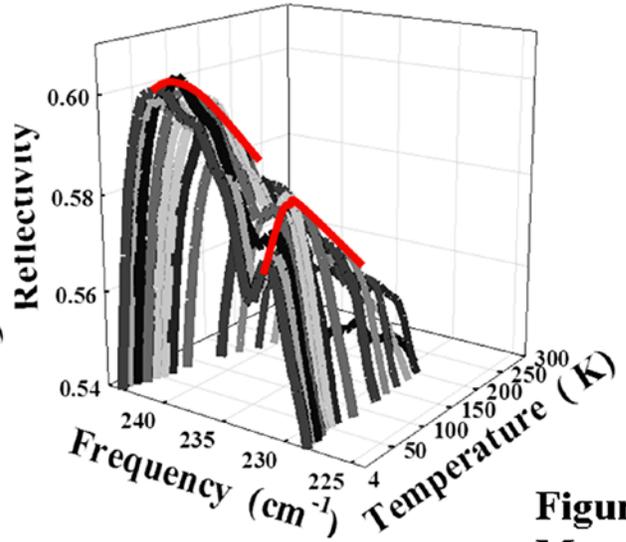

Figure 4
Massa et al

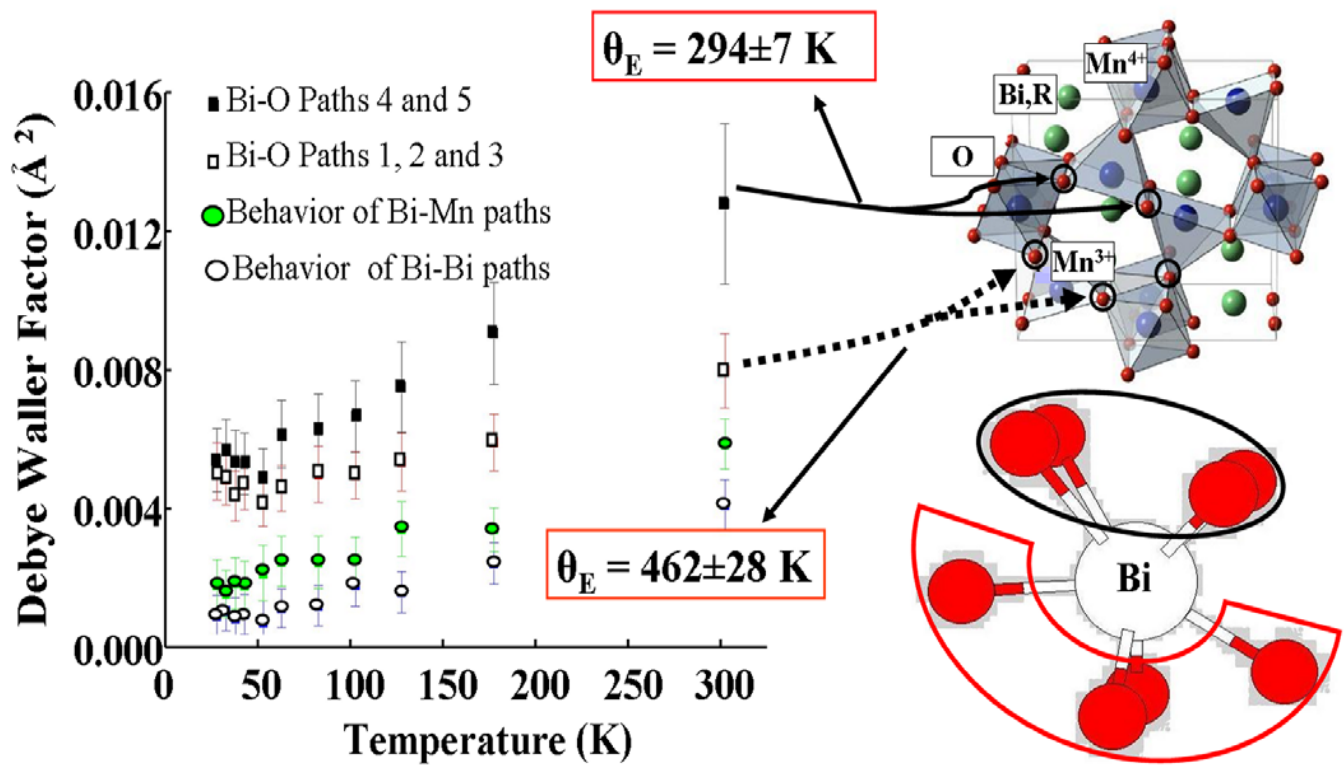

Figure 5
Massa et al

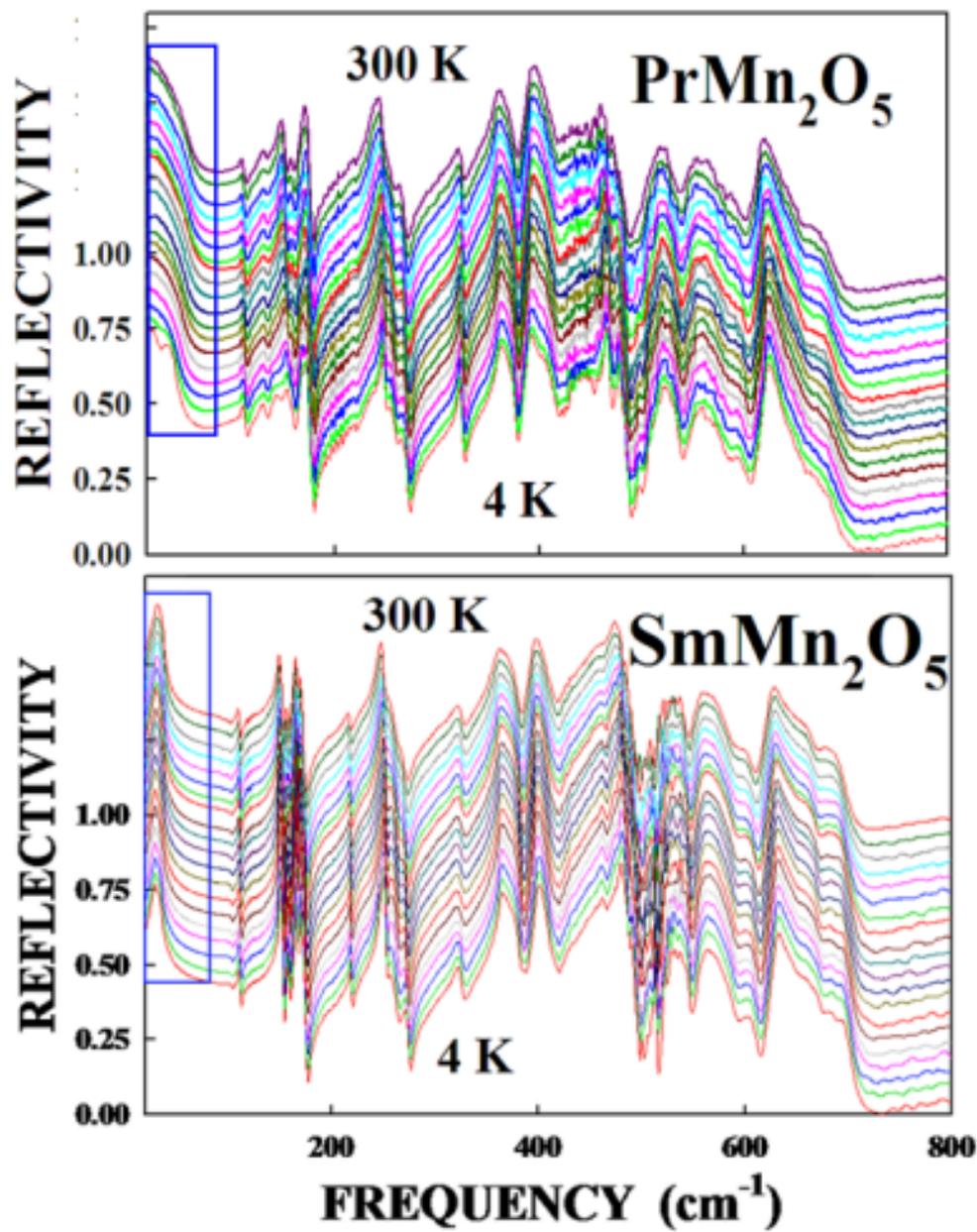

Fig. 6
Massa et al

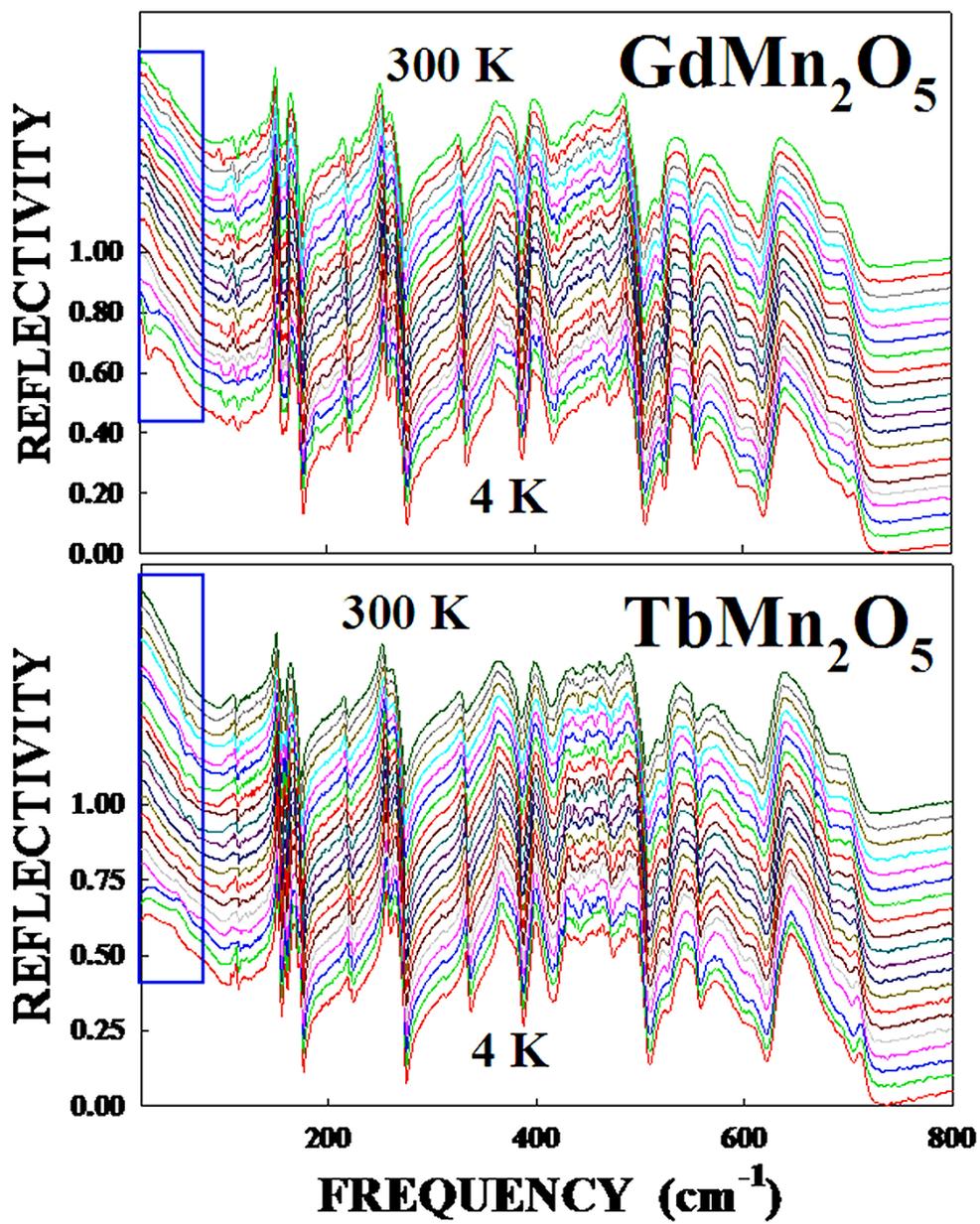

Figure 7
Massa et al

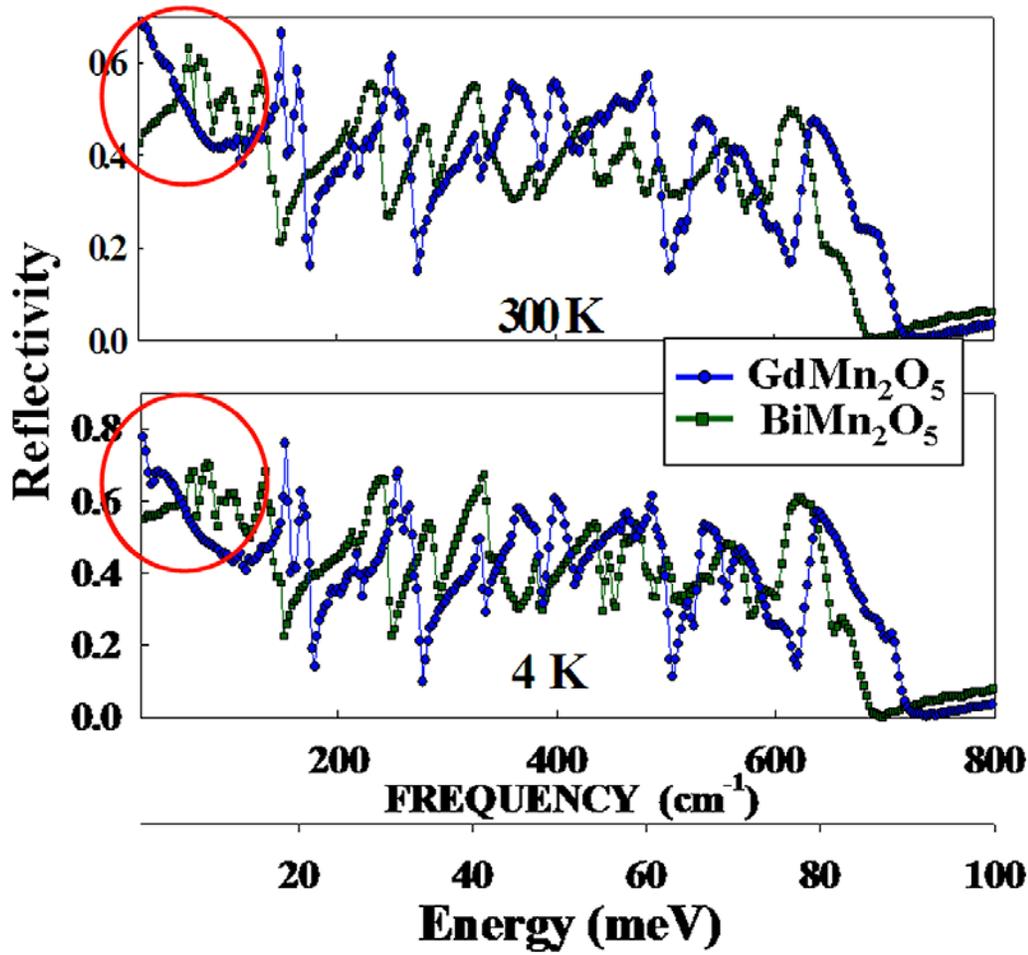

Figure 8
Massa et al

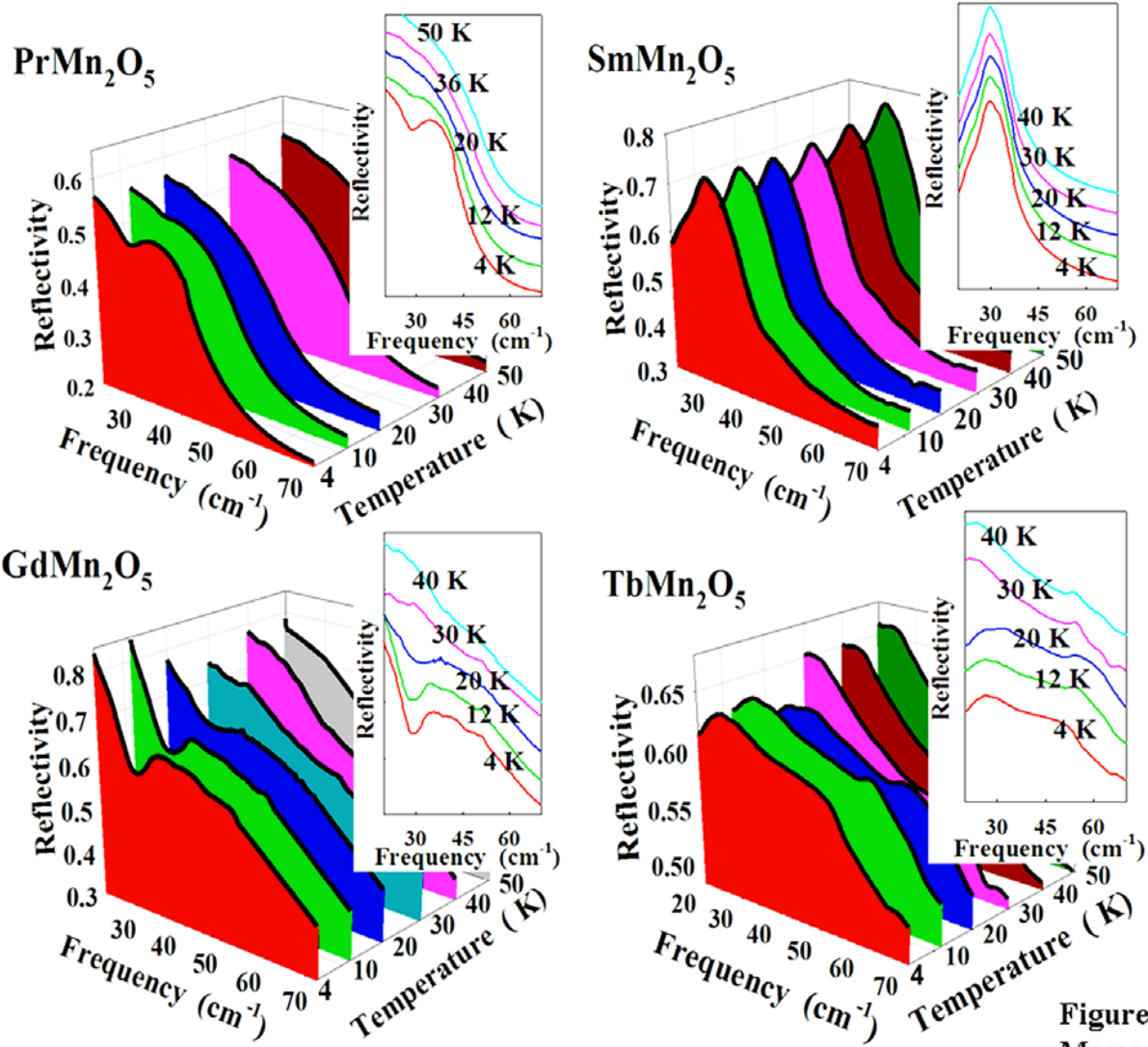

Figure 9
Massa et al

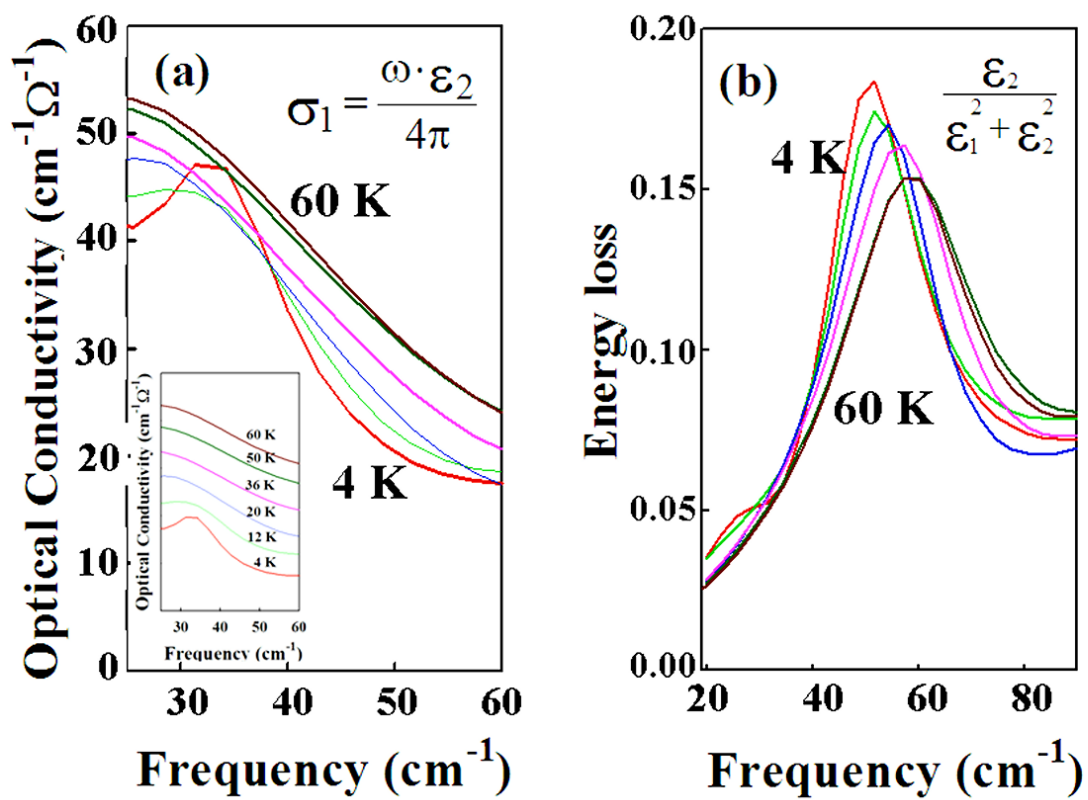

Figure 10
Massa et al

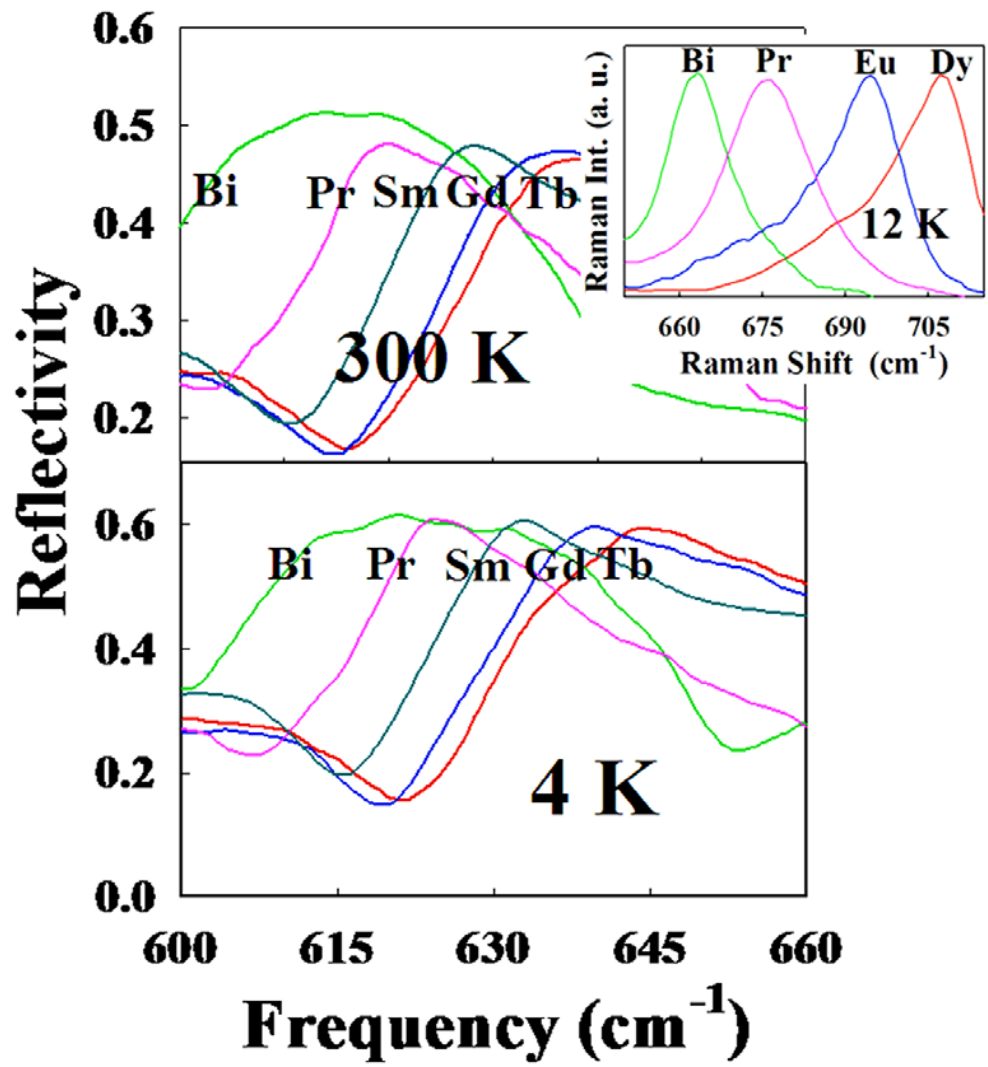

Figure 11
Massa et al